\documentclass[a4paper,fleqn,usenatbib]{mnras}

\usepackage{newtxtext,newtxmath}
\usepackage[T1]{fontenc}
\usepackage{ae,aecompl}

\usepackage{graphicx}	% Including figure files
\usepackage{amsmath}	% Advanced maths commands
\usepackage{amssymb}	% Extra maths symbols
\newcommand{\angstrom}{\textup{\AA}}

\usepackage{caption,booktabs}
\usepackage{cleveref}
\usepackage{tabularx}

\def\ra#1#2#3{#1$^{\rm h}$#2$^{\rm m}$#3$^{\rm s}$}
\def\dec#1#2#3{$#1^\circ#2'#3''$}

\title[PS1-13cbe as a ``Changing Look'' AGN]{PS1-13cbe: The Rapid ``Turn on'' of a Seyfert 1}

\author[Katebi et al.]{
Reza Katebi$^{1}$\thanks{E-mail: rk726014@ohio.edu},
Ryan Chornock$^{1}$\thanks{E-mail: chornock@ohio.edu},
Edo Berger$^{2}$, 
David O. Jones$^{3}$,
\newauthor
Ragnhild Lunnan$^{4}$,
Raffaella Margutti$^{5}$,
Armin Rest$^{6, 7}$,
Daniel M. Scolnic$^{8}$,
\newauthor
William S. Burgett$^{9}$,
Nick Kaiser$^{10}$,
Rolf-Peter Kudritzki$^{11, 12}$,
Eugene A. Magnier$^{12}$,
\newauthor
Richard J. Wainscoat$^{12}$,
Christopher Waters$^{12}$
\\
$^{1}$ Astrophysical Institute, Department of Physics and Astronomy, 251B Clippinger Lab, Ohio University, Athens, OH 45701, USA \\
$^{2}$ Harvard-Smithsonian Center for Astrophysics, 60 Garden Street, Cambridge, MA 02138, USA \\
$^{3}$ Department of Astronomy and Astrophysics, University of California, Santa Cruz, CA 95064, USA \\
$^{4}$ The Oskar Klein Centre \& Department of Astronomy, Stockholm University, AlbaNova, SE-106 91 Stockholm, Sweden \\
$^{5}$ Center for Interdisciplinary Exploration and Research in Astrophysics (CIERA) and Department of Physics and Astronomy,\\ Northwestern University, Evanston, IL 60208, USA \\
$^{6}$ Space Telescope Science Institute, 3700 San Martin Drive, Baltimore, MD 21218, USA \\
$^{7}$ Department of Physics and Astronomy, Johns Hopkins University, 3400 North Charles Street, Baltimore, MD 21218, USA \\
$^{8}$ Kavli Institute for Cosmological Physics, The University of Chicago, Chicago, IL 60637, USA \\
$^{9}$ GMTO Corp., 251 S. Lake Ave., Pasadena, CA 91101 USA \\
$^{10}$ Ecole Normale Suprieure ENS, 24, rue Lhomond, Paris, 75005, France \\
$^{11}$ University Observatory Munich, Scheinerstr. 1, D-81679 Munich, Germany\\
$^{12}$ Institute for Astronomy, University of Hawaii, 2680 Woodlawn Drive, Honolulu HI 96822, USA \\
}

\date{Accepted XXX. Received YYY; in original form ZZZ}

\pubyear{2017}

\begin{document}
\label{firstpage}
\pagerange{\pageref{firstpage}--\pageref{lastpage}}
\maketitle

% Abstract of the paper
\begin{abstract}
We present a nuclear transient event, PS1-13cbe, that was first discovered in the Pan-STARRS1 survey in 2013. The outburst occurred in the nucleus of the galaxy SDSS J222153.87+003054.2 at $z = 0.12355$, which was classified as a Seyfert 2 in a pre-outburst archival Sloan Digital Sky Survey (SDSS) spectrum. PS1-13cbe showed the appearance of strong broad H$\alpha$ and H$\beta$ emission lines and a non-stellar continuum in a Magellan spectrum taken 57 days after the peak of the outburst that resembled the characteristics of a Seyfert 1. These broad lines were not present in the SDSS spectrum taken a decade earlier and faded away within two years, as observed in several late-time MDM spectra. We argue that the dramatic appearance and disappearance of the broad lines and factor of $\sim 8$ increase in the optical continuum is most likely caused by variability in the pre-existing accretion disk than a tidal disruption event, supernova, or variable obscuration. The timescale for the turn-on of the optical emission of $\sim 70$ days observed in this transient is among the shortest observed in a ``changing look'' active galactic nucleus. 
\end{abstract}

% Select between one and six entries from the list of approved keywords.
% Don't make up new ones.

%%%%%%%%%%%%%%%%%%%%%%%%%%%%%%%%%%%%%%%%%%%%%%%%%%

%%%%%%%%%%%%%%%%% BODY OF PAPER %%%%%%%%%%%%%%%%%%
\section{Introduction}
The axisymmetric unification model of active galactic nuclei (AGN) suggests that there can be different types of AGNs based on the angle toward the line of sight. However, it is believed that the simple orientation-based model cannot fully explain the Type 1 (both narrow and broad emission lines are present in the spectra) and Type 2 (only narrow emission lines are present in the spectra) classification scheme because this model is challenged when different AGN types are observed in the same object at different epochs of time. One alternative suggestion is that some Type 2s were Type 1s and their engine is now not active anymore or basically turned off; therefore, the Type 2s are evolved version of the Type 1s \citep{penston1984evolutionary, runnoe2016now}. This alternative view was first suggested  based on the observations of broad emission lines disappearance in the Type 2 NGC 4151 \citep{lyutyj1984ngc,penston1984evolutionary} and 3C 390.3 \citep{penston1984evolutionary}. 

The term ``changing-look" (CL) was first used for the AGNs that showed X-ray absorption variations \citep{matt2003changing, bianchi2009complex, puccetti2007rapid, risaliti2009variable, marchese2012ngc}. Recently, this term has also been used to describe the type of AGNs that have been observed to optically transition from Type 1 to Type 1.8, 1.9 and 2 or vice versa, by the sudden appearance or disappearance of broad emission lines such as the broad H$\beta$ emission line \mbox{\citep{macleod2016systematic, runnoe2016now}}. 
In some objects, the broad emission lines disappeared and the continuum faded completely \citep{collin1973discussion, tohline1976variation, sanmartim20142d, denney2014typecasting, barth2015lick, runnoe2016now}, while in others broad emission lines appeared or, in other words, the AGN ``turned-on" \citep{cohen1986variability, storchi1993double, aretxaga1999seyfert, eracleous2001ngc, shappee2014man}. These objects are mostly at low redshift with low absolute luminosities. However, recently \cite{lamassa2015discovery} discovered a luminous changing look AGN with a redshift of $z=0.31$ that transitioned from Type 1 quasar to a Type 1.9 AGN in $\sim 9$ years. Subsequent studies \citep{macleod2016systematic,runnoe2016now,macleod2018} have started finding larger samples in the Sloan Digital Sky Survey (SDSS). Most recent, \cite{gezari2017iptf} discovered a quasar with a rapid ``turn-on" timescale of $< 1$ year that demonstrated one of the fastest changes of state to date.

The luminosities of Seyfert galaxies have been observed to vary over time. Variability has been detected in a wide range of wavelengths with timescales from hours to years with the shortest in X-rays that show variation time-scales of a few hours in Seyfert galaxies \citep{green1993nature,nandra1997asca}. The observed average variability time-scale in the optical in Seyfert galaxies is on the order of weeks to months \citep{kaspi1996multiwavelength,giveon1999long}. Accretion disk instabilities are the most promising model as the source of this variability \citep{kawaguchi1998optical, kelly2009variations}.

Here, we report the rapid turn-on ($\sim 70$ days) of a nuclear transient, PS1-13cbe, from a galaxy at redshift $z= 0.12355$ classified as a Seyfert 2 in a pre-event SDSS spectrum which was accompanied by the appearance of broad Balmer lines and hence a transition to a Seyfert 1.  This represents the most rapid ``turn-on" in a changing look AGN to date. We discuss the observations in Sections~\ref{Observations} and ~\ref{feature}. We discuss the possible scenarios for the origin of the variations observed in PS1-13cbe in Section~\ref{source}. We summarize and present our conclusions in Section~\ref{Conclusion}. Throughout this work we are assuming a standard $\Lambda$CDM cosmology with $H_{0} = 69.6$ km~s$^{-1}$~Mpc$^{-1}$, $\Omega_{m} = 0.286$ and $\Omega_{\Lambda} = 0.714$ parameters \citep{bennett20141} that yields a luminosity distance of $d_{L} =582$ Mpc. All magnitudes are in the AB system and all dates are UT. We assume a Galactic extinction value of $E(B-V)= 0.06$ \citep{schlafly2011measuring}.

%%%%%%%%%%%%%%%%%%%%%%%%%%%%%%%%%%%%%%%%%%%%%%%%%%

%%%%%%%%%%%%%%%%% Observation %%%%%%%%%%%%%%%%%%
\section{Observations of PS1-13cbe}\label{Observations}
The Medium Deep Survey (MDS) of the Pan-STARRS1 (PS1) sky survey performed daily (in-season) deep monitoring of ten $\sim$ 7 sq. deg. fields over the years 2010--2014 to find transient and variable sources.
The typical observation sequence was composed of $g_{\mathrm{P1}}$ and $r_{\mathrm{P1}}$ bands on the first night, $i_{\mathrm{P1}}$ on the next night, and then $z_{\mathrm{P1}}$ on the third night. This pattern was repeated during the $\sim$ 6 months of the observing season and was only interrupted by the weather and times near full moon, when observations in the $y_{\mathrm{P1}}$ filter were taken. A more complete description of the survey and photometric system were given by \citet{tonry2012pan} and \citet{chambers2016pan}.
%The MDS is ideal for detecting transients such as supernovae (SNe). 

On 2013 July 9, we detected a transient event, PS1-13cbe, at coordinates $\alpha=$ \ra{22}{21}{53.86}, $\delta=$ \dec{+00}{30}{54.56} (J2000) coincident with the nucleus of a galaxy using the \texttt{photpipe} transient discovery pipeline, described by \citet{rest2014cosmological} and \citet{scolnic2018complete}. The host galaxy was observed as part of SDSS and given the name SDSS J222153.87+003054.2 (hereafter SDSS J2221+0030), with a spectroscopic redshift of $z=0.12355$ \citep{ahn2014tenth}. We constructed a difference image light curve for PS1-13cbe using the PS1 transient pipeline \citep{scolnic2018complete}.  The template images were created from a stack of high-quality observations excluding the observing season containing the outburst (the year 2013) and were then subtracted from all observations of the transient.  It is important to note that this photometry represents a flux difference relative to the host contribution present in the template (which is consistent with the SDSS photometry; \citealt{ahn2014tenth}). We include the PS1 photometry in Table~\ref{phot_tab}.

\subsection{Optical photometry}
\begin{figure}
	\includegraphics[width=\columnwidth, scale = 0.98]{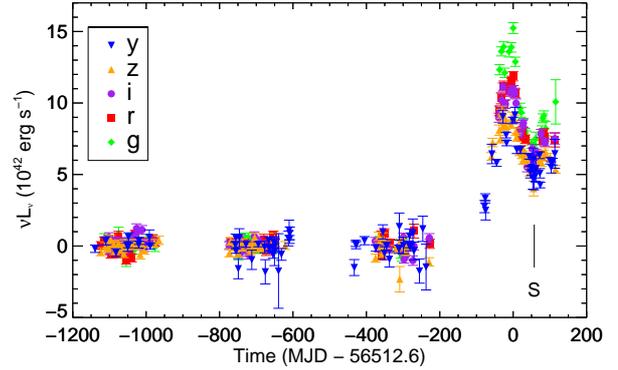}
	\caption{The observed transient luminosities of PS1-13cbe from the PS1 survey in $grizy_{\mathrm{P1}}$ filters after correction for Galactic extinction. 
	S: marks the epoch of the LDSS spectrum (MJD 56570).}
	\label{PS1_Flux}
\end{figure}

\begin{figure}
	\includegraphics[width=\columnwidth , scale = 0.98]{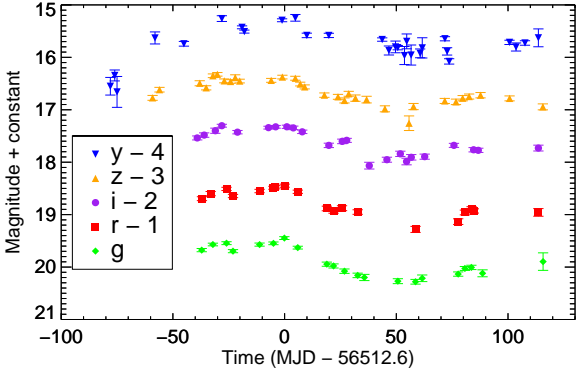}
	\caption{Light curves of PS1-13cbe in $grizy_{\mathrm{P1}}$ bands, corrected for Galactic extinction. Note the small-scale fluctuations in the light curve (such as near $-20$~d).}
	\label{PS1_lightCurve}
\end{figure}

As shown in Figure~\ref{PS1_Flux}, the light curves of PS1-13cbe were constant and consistent with zero change in flux relative to the baseline in the template for three observational seasons and then showed a rise peaking at MJD 56512.6, decline, and a second rise ($g_{\mathrm{P1}} = 19.5$ mag at peak; see Figure~\ref{PS1_lightCurve}). The $g_{\mathrm{P1}}$ band luminosity showed a rise of about $\sim 1.5 \times 10^{43}$ erg~s$^{-1}$ from the base luminosity in the course of $\sim 70$ days and then declined to $\sim 0.7 \times 10^{43}$ erg~s$^{-1}$ in $\sim 50$ days and rose back up again to $\sim 1 \times 10^{43}$ erg~s$^{-1}$ in the course of next $\sim 50$ days, at which point the MDS ended.


With the SDSS value for the quiescent host flux, $g^{\prime}= 19.21 \pm 0.01$ mag  \citep{ahn2014tenth}, and $g_{\mathrm{P1}} = 19.5$ mag for PS1-13cbe at the peak of the outburst, we can see that the total luminosity of the galaxy  increased by $\gtrsim$ 75\%.  However, the quiescent host flux value is dominated by star light. We estimate that the central AGN contributes $\lesssim$10\% of the continuum flux in the quiescent spectrum taken by SDSS because the absorption lines from star light are not noticeably diluted by a non-stellar continuum, so the amplitude of the outburst from the AGN must be significantly larger, as discussed below.

After the MDS ended, we obtained late-time photometry in $g^{\prime}$ using the MDM4K and Templeton detectors on the 2.4 m Hiltner and 1.3 m McGraw-Hill telescopes at MDM Observatory on 2015 November 18 and 2017 June 18, respectively. The total magnitude of the host galaxy (including any possible transient contribution) was $g^{\prime} = 19.23 \pm 0.05$~mag ($g^{\prime} = 19.22 \pm 0.13 $ mag) at $831$ ($1408$) days after the peak, which are consistent with the SDSS pre-outburst photometry.  Therefore, the system returned to the baseline flux value in $\lesssim 2$~yr after the outburst.

\subsection{Observations of the host galaxy}
In addition to the SDSS observation of the host ($\sim10$ years before the outburst), the source was detected by the AllWISE survey in the \textit{W1, W2, W3,} and \textit{W4} bands with cataloged values of $17.82 \pm 0.037$, $18.20 \pm 0.34$, $16.65 \pm 0.23$ and $15.13 \pm 0.37$ magnitudes, respectively \citep[$\sim 3$ years before the outburst;][]{chang2015stellar}. We also obtained photometry using the Neil Gehrels \textit{Swift} Observatory \citep{gehrels2004swift} with the UV Optical Telescope \citep[UVOT;][]{roming2005swift} in the \textit{u, uvw1,} and \textit{uvw2} filters on 2016 November 27--29 ($\sim 3$ years after the outburst) and we measure values of $21.10 \pm 0.13$, $21.76 \pm 0.21$ and $22.41 \pm 0.19$ magnitudes, respectively. 
%of $21.096 \pm 0.126$, $21.758 \pm 0.206$ and $22.407 \pm 0.189$ magnitudes, respectively.

\subsection{X-ray photometry}\label{X-ray}
We obtained an X-ray observation of SDSS J2221+0030 using the X-ray telescope \citep[XRT;][]{burrows2003swift} on \textit{Swift} with a total exposure time of $7.1$ ks between 2016 November 27 and November 29. The host was not detected with a $3\sigma$ upper limit of $2.13 \times 10^{-3}$ cts~s$^{-1}$ ($0.3-10$ keV). Using the Galactic neutral hydrogen column density in the direction of PS1-13cbe of $N_{H} = 4.49 \times 10^{20}$ cm\textsuperscript{-2} \citep{kalberla2005leiden} and assuming no intrinsic absorption, and a typical photon index of $\Gamma = 2$, we calculate the unabsorbed X-ray flux to be $f_{x}(2-10$ keV$) \le 4.0\times10^{-14}$ erg~cm$^{-2}$~s$^{-1}$, which translates to $L_{X} \le 1.6 \times 10^{42}$ erg~s$^{-1}$. 

The absorbing column densities of Seyfert 2 galaxies range from $10^{22}$ to $10^{25}$ cm\textsuperscript{-2} (e.g., \citealt{risaliti1999distribution}) and therefore, assuming a minimum intrinsic column density of $10^{22}$ cm\textsuperscript{-2} for the host galaxy in addition to the Galactic value, we recalculate the unabsorbed X-ray flux to be $f_{x}(2-10$ keV$) \le 1.0\times10^{-13}$ erg~cm$^{-2}$~s$^{-1}$ or $L_{X} \le 4.1 \times 10^{42}$ erg~s$^{-1}$. Additionally, using $L_{X}$ and the empirical bolometric correction from \cite{marconi2004local}, this corresponds to a minimal upper limit on the bolometric luminosity of the object $L_{\rm bol} \le 0.6 \times 10^{44}$ erg~s$^{-1}$ 3 years after the outburst, although this value can increase if the intrinsic absorption is higher (e.g., $L_{\rm bol} \le 9.0 \times 10^{45}$ erg~s$^{-1}$ for an intrinsic column density of $10^{24}$ cm\textsuperscript{-2}). We did not find any archival X-ray observations of the object before or at the time of the outburst. 

\subsection{Optical spectroscopy}\label{Spectroscopy}
A pre-outburst spectrum of SDSS J2221+0030 was obtained on 2003 May 26 by SDSS \citep{ahn2014tenth}. After the detection by PS1, we observed PS1-13cbe on 2013 October 5 (+57 days after peak) for $1200$ s using the Low Dispersion Survey Spectrograph-3 (LDSS3) on the 6.5 m Magellan Clay telescope. We used a 1$\arcsec$-wide long slit with the VPH-all grism to cover the observed wavelength range $3700-10000$~\AA\ with a resolution of $\sim$ 9~\AA.

At late times, we obtained five epochs of spectroscopy using the Ohio State Multiple Object Spectrograph on the 2.4 m Hiltner telescope at MDM Observatory \citep[OSMOS; ][]{martini2011ohio}. The first two spectra were taken on 2015 October 3 and 2015 November 12 using the $1.2\arcsec$ center slit with a VPH-red grism and an OG530 filter that covered $5350-10200\, \angstrom$ (resolution = $5 \, \angstrom$) in the observed frame. No significant spectral differences were present, so we combined these two observations for all subsequent analysis. The next two spectra were obtained on 2015 December 1 and 2016 November 16 using the same $1.2\arcsec$ center slit and a VPH-blue grism that covered $3675-5945 \, \angstrom$ (resolution = $2 \, \angstrom$) in the observed frame. The final spectrum was taken on 2017 June 18 using a $1.0\arcsec$ outer slit and VPH-red grism that covered $3930-9050 \, \angstrom$ (resolution = $4\,\angstrom$) in the observed frame.

\begin{figure*}
	\includegraphics[scale=0.8]{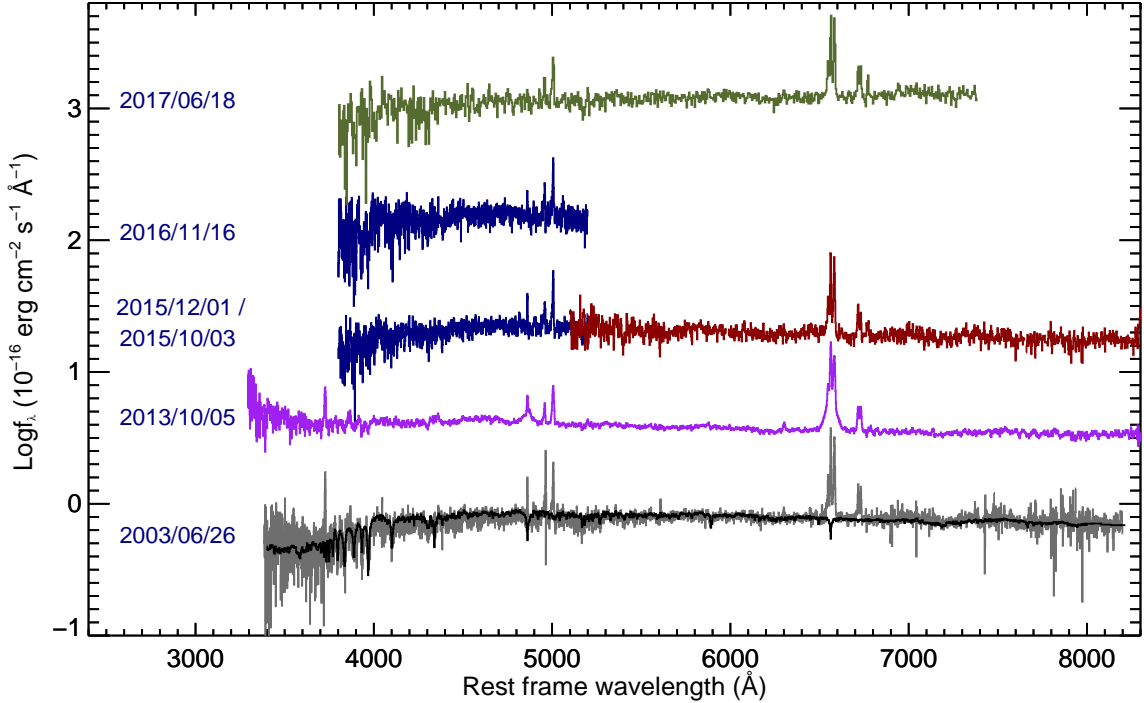}
	\caption{Optical spectra for PS1-13cbe, from bottom to top: host galaxy of PS1-13cbe from SDSS (grey), model of the host galaxy generated by FAST 1.0 (black), spectrum during outburst observed using LDSS3 (purple), spectra obtained with OSMOS in red and blue for the red and blue setups, respectively, and the most recent OSMOS spectrum (olive green).}
	\label{PS1_spectra}
\end{figure*}

We preprocess our data using standard procedures such as flat-fielding, bias subtraction, and wavelength calibration using arc lamps in IRAF \footnote{IRAF is distributed by the National Optical Astronomy Observatories, which are operated by the Association of Universities for Research in Astronomy, Inc. (AURA) under cooperative agreement with the National Science Foundation.}. Additionally, we removed cosmic rays using the L.A.Cosmic \citep{van2001cosmic} task. Finally, we calibrate our data using our own IDL procedures and observations of the standard stars  BD+174708 for red spectra and Feige110 for blue spectra. All of the spectra for PS1-13cbe are shown in Figure~\ref{PS1_spectra}.

\section{Observational features of PS1-13cbe}\label{feature}
\subsection{Host galaxy of PS1-13cbe}\label{Host-Galaxy}
To isolate the emission lines of SDSS~J2221+0030 and correct for stellar absorption lines, we simulated a galaxy model with the FAST 1.0 code \citep{kriek2009ultra} using the archival SDSS spectrum and optical photometry in \textit{ugriz} bands along with the \textit{Swift}  \textit{u}, \textit{uvw1}, and \textit{uvw2} photometry. We experimented with the initial parameters and generated the best fit with the stellar age of $3 \times 10^{9}$ years and e-folding timescale of $\tau \approx 10^{9}$ years, by assuming the star-formation history to be exponentially declining, the stellar initial mass function from \citet{chabrier2003galactic}, the \cite{bruzual2003stellar} spectral library, a Milky Way dust law \citep{cardelli1989relationship}, and solar-like metallicity of $Z = 0.02$. The FAST model of the host galaxy is over-plotted on the original SDSS spectrum (grey) in the bottom of the Figure~\ref{PS1_spectra} (black).

\begin{figure*}
	\includegraphics[scale =0.8]{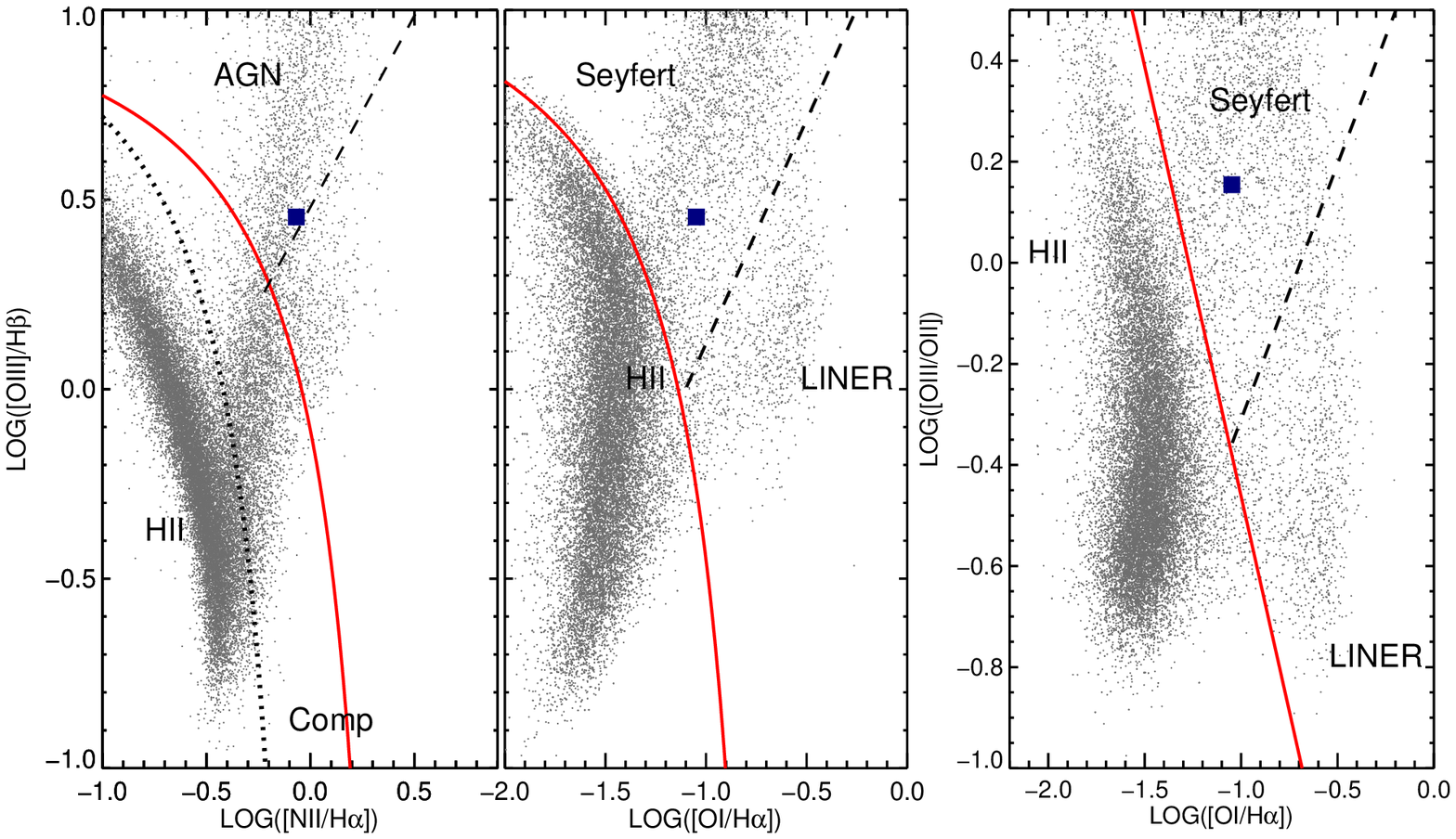}
	\caption{The excitation diagrams using the [N II]$\lambda6583/H\alpha$, [O I]$\lambda6300/H	\alpha$, [O III]$\lambda5007/H\beta$ and [O III]$\lambda5007/$[O II]$\lambda3726$ line ratios \citep{baldwin1981classification,kewley2006host}. The navy blue squares display the position of the host galaxy of PS1-13cbe. The shaded area represents the location of the SDSS galaxies calculated using \href{http://	wwwmpa.mpa-garching.mpg.de/SDSS/DR7/}{MPA-JHU} where darker regions represent higher number density of the galaxies. The extreme star formation line \citep[solid red; ][]{kewley2001theoretical, kewley2006host}, the revised star formation line \citep[dotted black; ][]{kauffmann2003host} and the Seyfert-LINER classification line \citep[dashed black; ][]{kewley2006host,fernandes2010alternative} are also plotted. Comp: AGN/star forming composites.}
	\label{Host_LineRatio}
\end{figure*}

After subtracting this model from the pre-outburst optical spectrum of the host galaxy of PS1-13cbe  (bottom of Figure~\ref{PS1_spectra} in grey), we fit the profiles of the narrow emission lines [S II]$\lambda\lambda6717,6731$, [O III]$\lambda\lambda4959,5007$, [O II]$\lambda3727$, [O I]$\lambda6300$,  [N II]$\lambda\lambda6549,6583$, H$\alpha$, and H$\beta$. It has been shown that a model for narrow emission lines can be obtained from [S II] lines in most cases \citep{ho1997search}. By using this fact, we constrained parameters of the Gaussian profiles of the other narrow lines. The lines were not well modeled with single Gaussians, so we used double Gaussian profiles to fit the line profile of all the narrow lines except [O II]$\lambda3726$ and [O I]$\lambda6300$. In Figure~\ref{Host_LineRatio}, we plot the line ratios in excitation diagrams \citep{baldwin1981classification}. We also show the extreme star formation line \citep{kewley2001theoretical, kewley2006host}, the pure star formation line \citep{kauffmann2003host} and the Seyfert-LINER classification line \citep{kewley2006host,fernandes2010alternative}. Additionally, we show 30000 randomly selected galaxies (shaded area in Figure~\ref{Host_LineRatio}) with emission-line fluxes from the MPA-JHU DR7 catalog \footnote{\href{http://wwwmpa.mpa-garching.mpg.de/SDSS/DR7/}{http://wwwmpa.mpa-garching.mpg.de/SDSS/DR7/}} \citep{aihara2011eighth}. The automatic Portsmouth pipeline from SDSS classified the host galaxy of PS1-13cbe as a LINER \citep{sarzi2006sauron}; however, our emission line ratios calculated from the SDSS spectrum after subtraction of the stellar continuum (navy squares) classify the host galaxy as a clear Seyfert (Figure~\ref{Host_LineRatio}).

\begin{figure*}
	\includegraphics[scale = 0.8]{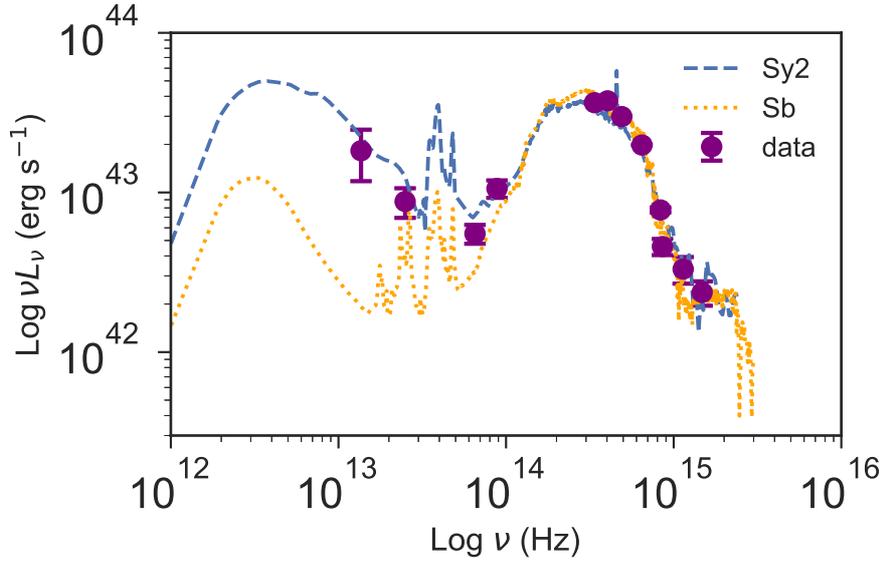}
	\caption{SED of SDSS J2221+0030 in quiescence (purple circles), scaled SED templates of an Sb spiral galaxy (dotted orange line) and a Seyfert 2 galaxy (dashed blue lines) from the SWIRE template library \citep{polletta2007spectral}. The host exhibits a mid-infrared (MIR) excess relative to the star-forming template due to an AGN.}
	\label{SED}
\end{figure*}

Next, we constructed the spectral energy distribution (SED) of the host galaxy using the observed and archival photometry data. We also scaled and plotted SEDs of an Sb spiral galaxy and a Seyfert 2 galaxy from the SWIRE template library \citep{polletta2007spectral} alongside the SED of the host (see Figure~\ref{SED}). The SED of the Seyfert 2 galaxy fits well from the mid-IR to UV, including the contribution from star light. As we can see in Figure~\ref{SED}, the SED of the spiral galaxy fits the optical and UV part of the host SED as well; however, there is a broad excess in the mid-IR (\textit{W3} and \textit{W4}). This excess is because of the contribution of dust heated by the AGN in the host galaxy. After subtracting the host contribution (using the Sb template) from the observed values, we estimate the luminosity of the AGN component to be $\nu L_{\nu}(W4) \approx (1.63 \pm 0.65) \times 10^{43}$ erg~s$^{-1}$. Then, using a bolometric correction factor of $10.1 \pm 1.4$ for  the \textit{W4} band from \cite{runnoe2012updating}, we estimate the total bolometric luminosity of the AGN to be $L_{\mathrm{bol}}(W4) \approx  (1.65 \pm 0.65) \times 10^{44}$ erg~s$^{-1}$.

Furthermore, narrow emission lines are possible indicators of the intrinsic bolometric luminosity \citep{netzer2009accretion}. We estimate the bolometric luminosity of the AGN from the luminosity of the narrow [O III]~$\lambda$5007 emission line in the SDSS spectrum and used the conversion that $L_{\mathrm{bol}} = 3500 L$($\lambda$5007), which has a variance of 0.38 dex by assuming a standard AGN SED \citep{heckman2004present}. This results in an estimate of $L_{\mathrm{bol}} = 1.6 \times 10^{44}$ erg~s$^{-1}$, which is consistent with the bolometric luminosity calculated  using the \textit{W4} band above. This luminosity can be an overestimate if some of the [\ion{O}{III}] emission originates from star formation. However, as shown in Figure~\ref{Host_LineRatio}, SDSS J2221+0030 is classified as a clear Seyfert galaxy and therefore narrow line excitation is dominated by the AGN. Moreover, as we discussed in Section~\ref{X-ray}, we estimate the upper limit on the bolometric luminosity in quiescence from the X-ray non-detection to be  $\le 0.6 \times 10^{44}$ erg~s$^{-1}$, which is consistent with the bolometric luminosity estimated before the outburst in quiescence. This is only an estimate because the X-ray flux is a non-detection, whereas having an X-ray detection would help us to have a better estimate of the intrinsic absorption and column density.  

\subsection{Astrometry}
\begin{figure}
	\includegraphics[width=\columnwidth , scale = 0.5]{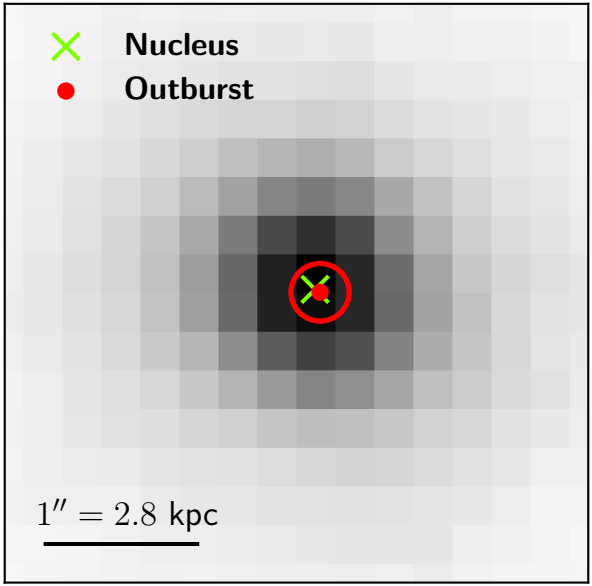}
	\caption{The $g_{\mathrm{P1}}$ PS1 template image of  SDSS J2221+0030  showing the position of the centroid of the galaxy (green ``X"), the position of PS1-13cbe (red dot), and its position uncertainty is shown by a circle with $5\sigma$ radius (red circle). The position uncertainty is dominated by the outburst.}
	\label{Astrometry}
\end{figure}

To find the location of PS1-13cbe in its host galaxy, we performed relative astrometry between the PS1/MDS template images of the host galaxy and the position of the transient reported by \texttt{photpipe}. First, we fit a 2-dimensional Gaussian function to the templates in all filters to find the centroid of SDSS J2221+0030 and then, using the weighted average centroid coordinates of the transient, we calculate the offset of PS1-13cbe from the center of its host galaxy to be $0.036\arcsec \pm 0.035\arcsec$ ($101 \pm 100$ pc), consistent with the nucleus. We used a systematic astrometric error floor of 0.1 pixel to calculate the uncertainty in the offset \citep{scolnic2018complete}. We show the position of PS1-13cbe relative to the center of its host in $g_{\mathrm{P1}}$ band in Figure~\ref{Astrometry}.

\subsection{Multi-band light curves of PS1-13cbe}
\begin{figure}
	\includegraphics[width=\columnwidth, scale = 1.2]{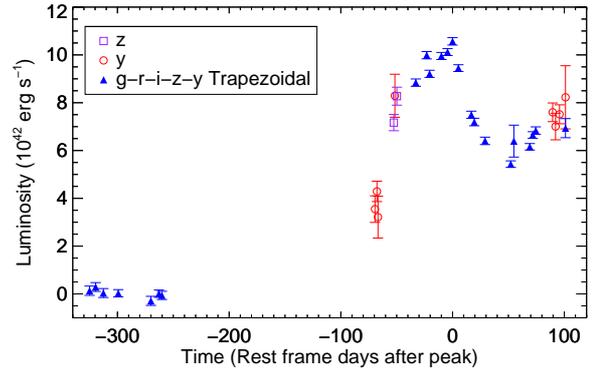}
	\caption{Total optical luminosity light curve of PS1-13cbe integrated over the $grizy_{\mathrm{P1}}$ filters and relative to the baseline flux in the template images. The total optical luminosity was estimated using the spectral distribution at each epoch and the trapezoidal rule (blue triangles). The estimated luminosity at early and late times using $y_{\mathrm{P1}}$ (red circles) and $z_{\mathrm{P1}}$ (purple squares) bands assumed the same colour correction as measured from those epochs with all filters.} 
 	\label{LuminosityVsTime}
\end{figure}

We calculate the total optical luminosity light curve of PS1-13cbe using multi-band observations shown in Figure~\ref{PS1_Flux}. This has been done by first finding the epochs where $g_{\mathrm{P1}}$ and $r_{\mathrm{P1}}$ bands were observed simultaneously and then interpolating the $i_{\mathrm{P1}}$, $z_{\mathrm{P1}}$, and $y_{\mathrm{P1}}$ bands at these epochs using Legendre polynomials. Next, by integrating the spectral distribution at each epoch and using the trapezoidal rule, we calculate the total flux and thereafter optical luminosity over $3685-8910$~\AA\ in the rest frame. At late and early times, we had data only from the $y_{\mathrm{P1}}$ and $z_{\mathrm{P1}}$ bands, so we calculate the luminosity assuming the same colour correction as measured from those epochs with all filters. As seen in Figure~\ref{LuminosityVsTime}, the total optical luminosity of PS1-13cbe rises to a peak value of $(1.06 \pm 0.01) \times 10^{43}$ erg~s$^{-1}$ in the course of $\sim 70$ days, declines in next $\sim 50$ days, and then rises back up again.  

\begin{figure*}
	\includegraphics[scale = 0.7]{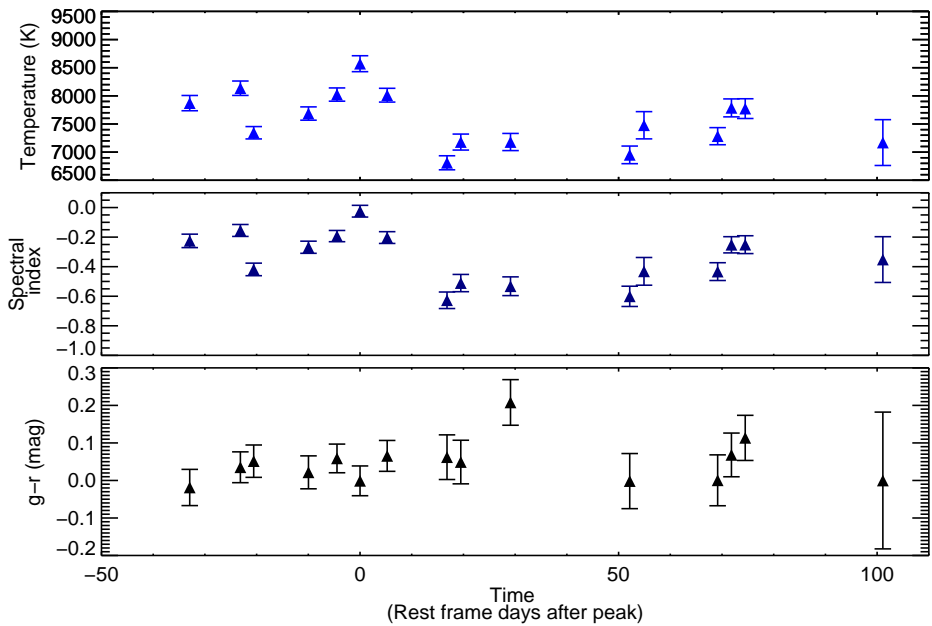}
	\caption{\textit{Top:} Rest-frame blackbody temperature from fitting the optical photometry. \textit{Middle:} Rest-frame spectral index from fitting a power law $f_{\nu}\propto \nu^{\alpha}$ to the optical photometry, where $\alpha$ is the spectral index. 
	\textit{Bottom:} observed \textit{g-r} colour diagram. All three panels show little or no evolution during the outburst.}
	\label{T_BB}
\end{figure*}

Furthermore, to study the evolution in continuum colour and temperature while remaining agnostic about the overall SED of the transient, we fit both power law and blackbody models. We also calculate the $g-r$ colour and show it alongside the spectral index and blackbody temperatures for all epochs in Figure~\ref{T_BB}. The reason we chose $g_{\mathrm{P1}}$ and $r_{\mathrm{P1}}$ bands as a proxy for the colour is that they were taken on the same night in the PS1 survey and do not need to be interpolated. No strong colour evolution during the outburst is evident in Figure~\ref{T_BB}.

\subsection{Spectral features of PS1-13cbe}
In Figure~\ref{PS1_spectra}, we show the optical spectra for PS1-13cbe. The spectra of PS1-13cbe show remarkable evolution over the course of fourteen years. The spectra contain narrow and broad emission line profiles, including hydrogen Balmer lines, [S II]$\lambda\lambda6717,6731$, [O III]$\lambda\lambda4959,5007$, [O II]$\lambda3726$, [O I]$\lambda6300$ and [N II]$\lambda\lambda6549,6583$. We note that the spectra were taken with different effective apertures (slit and seeing) and different spectral resolutions. We fit the H$\alpha$+[N II] and H$\beta$+[O III] complexes in order to investigate the presence of broad H$\alpha$ and H$\beta$ lines. 

First, we scaled our OSMOS (blue setup) and LDSS3 spectra using the flux of [O III] $\lambda 5007$ from the SDSS spectrum and scaled the OSMOS red setup using the [N II] flux from SDSS. We did these scalings with the assumptions that these narrow lines are centrally concentrated in the host and that the fluxes do not change in such a short timescale because it has been shown that narrow emission lines only slowly vary over decades \citep{peterson2013size}. In addition, the LDSS3 spectrum was taken during the outburst and contains transient flux. In that case, first we fit a power law $f_{\nu} \propto \nu^{\alpha}$ to the optical photometry from PS1 at the epoch when the spectrum was taken, where $\alpha = -0.58$. After that, we constructed a model from a linear combination of the power-law continuum and the host galaxy model shown in Figure~\ref{PS1_spectra}. Additionally, we smoothed this model by a Gaussian with full width at half maximum (FWHM) $= 5 \, \angstrom$ to better match the resolution of our data and subtracted it from the LDSS3 spectrum, isolating the emission-line spectrum.

Next, we fit the [S II] lines using double Gaussian profiles and use this model to constrain the multi-component Gaussian profiles that were used to fit the H$\alpha$+[N II] complex to reduce the number of free parameters \citep{ho1997search}.  
Specifically, we model emission lines using two Gaussian profiles for each narrow line and one broad component for H$\alpha$ and simultaneously fit for parameters of the broad and and narrow lines. We fix the narrow components to the wavelengths of H$\alpha$ and [N II] $\lambda\lambda6549,6583$. Also, we fix the widths and relative amplitudes of the two Gaussian components of the narrow lines using the values from the [S II] model, leaving only the overall amplitudes of the narrow H$\alpha$ and [\ion{N}{II}] lines as free parameters. The parameters of the broad component of H$\alpha$ were allowed to vary freely. We fit the [O III] $\lambda\lambda4959,5007$ lines using two Gaussian profiles for each narrow line without constraining the parameters. For H$\beta$, however, we use two Gaussian profiles, a single one for the narrow component with just the centroid fixed and allowing the other parameters to vary, and one for the broad component with no constraints.

\begin{figure*}
	\includegraphics[scale = 0.8]{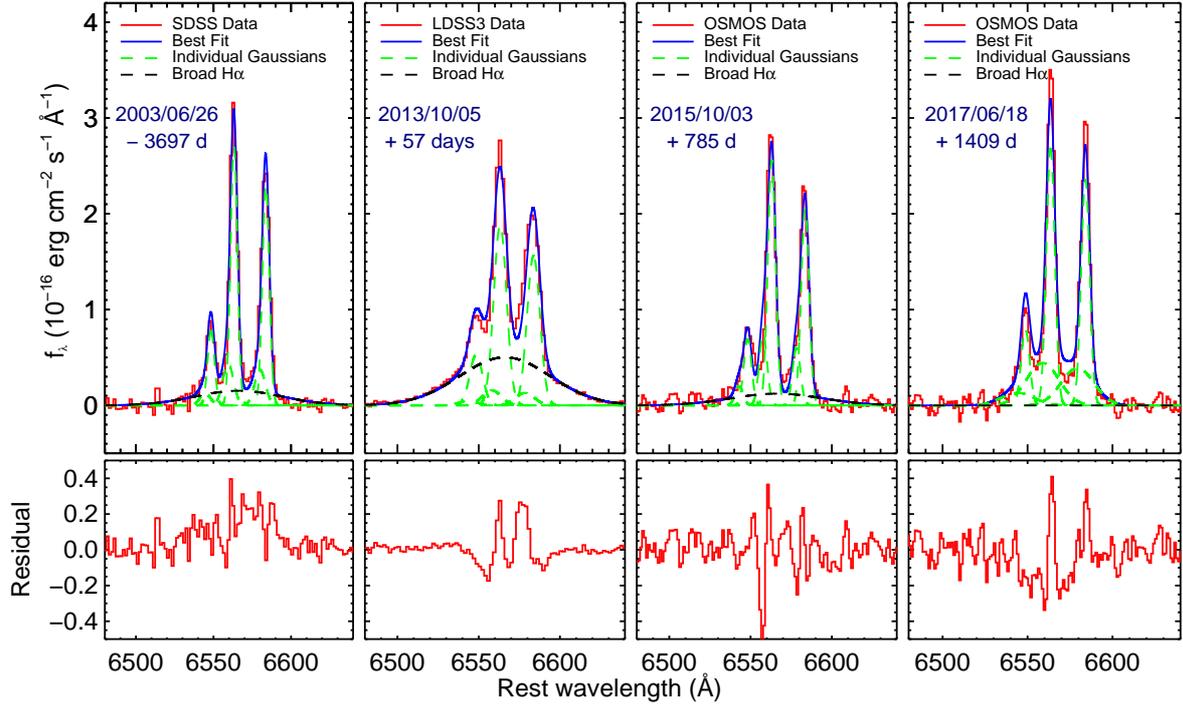}
	\caption{Continuum subtracted H$\alpha$ line profiles. \textit{Top:} Multiple-component Gaussian fit to the H$\alpha$+[NII] emission lines (blue), individual  components (dashed green) and broad component of the H$\alpha$ (dashed black). \textit{Bottom:} The fit residuals. \textit{Top-left (on each panel):} Observation date and numbers of days before/after the peak (navy blue).}
	\label{Halpha}
\end{figure*}

\begin{figure*}
	\includegraphics[scale = 0.8]{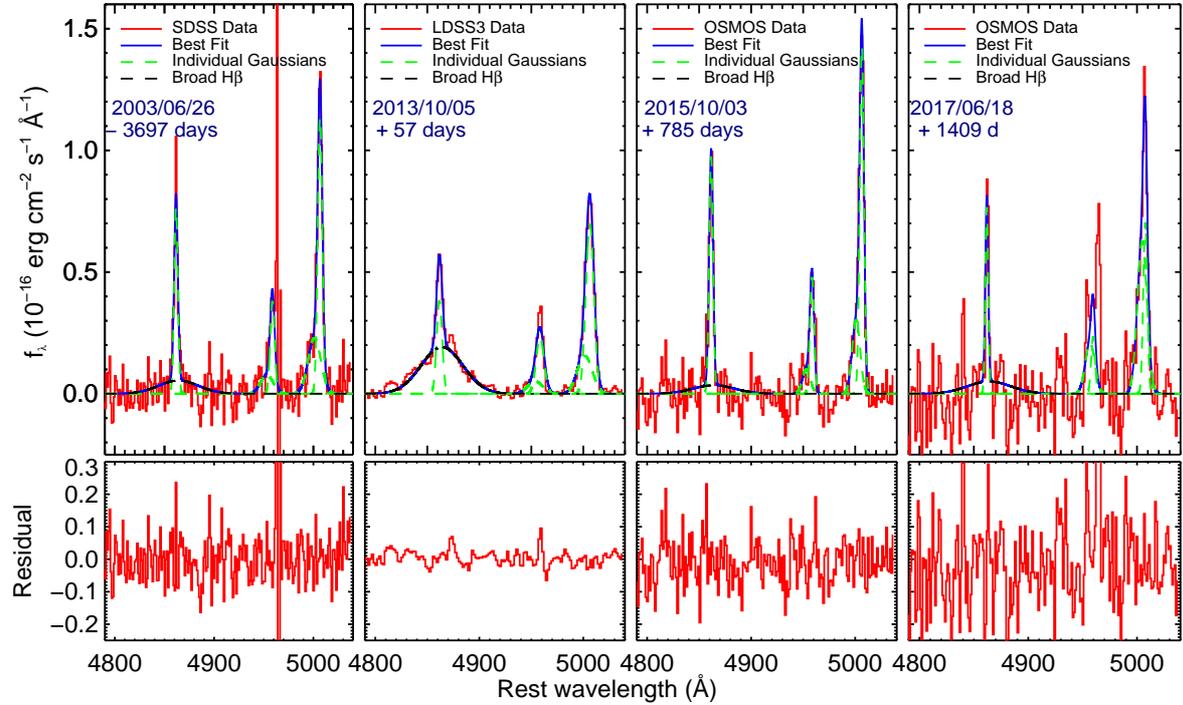}
	\caption{Continuum subtracted H$\beta$ line profiles. \textit{Top:} Multiple-component Gaussian fit to the H$\beta$+[O III] emission lines (blue), individual components (dashed green) and broad component of the H$\beta$ (dashed black). \textit{Bottom:} The fit residuals. \textit{Top-left (on each panel):} Observation date and numbers of days before/after the peak (navy blue). Strong residuals are present in the red wing of $\lambda4959$ due to poor subtraction of the $5577 \angstrom$ sky line.}
	\label{Hbeta}
\end{figure*}

The SDSS and OSMOS spectra lack transient flux, so we only subtract a scaled galaxy model and perform the same procedure as for the LDSS3 spectrum to fit the narrow and broad emission lines, with the exception of fixing the width and the the centroid of the broad component of both H$\alpha$ and H$\beta$ to the values derived from LDSS3 spectra, allowing only the normalization to vary. We show the resultant Gaussian fits to the H$\alpha$+[NII] and H$\beta$+[O III] emission lines in Figure~\ref{Halpha} and ~\ref{Hbeta} respectively. 

As we show in Figure~\ref{Halpha} and Figure~ \ref{Hbeta}, the existence of the broad H$\alpha$ and H$\beta$ components is clear in the LDSS3 spectrum (a weak broad H$\gamma$ emission line was also detected). \cite{ho1997search} found that they could  reliably extract a weak broad H$\alpha$ component that comprised $\ge$ 20\% of the H$\alpha$ + [NII] blend from spectra with spectral resolution close to ours. As shown in Figure~\ref{Halpha}, there is marginal evidence for a broad component of H$\alpha$ in the SDSS and OSMOS data, which is 21\% and 17\% of the flux in the H$\alpha$ + [NII] complex, respectively. However, it is notable that there are visible wiggles in the residuals of the fit from the spectra that suggest the broad emission line profiles are not purely Gaussian. For this reason, we are not confident that broad H$\beta$ line in the SDSS data and broad lines in the OSMOS are real and are not result of deviations from a Gaussian profile. We report the measured luminosities of the broad components of the H$\alpha$ and H$\beta$ emission lines in Table~\ref{table:2}. 

As mentioned in Section~\ref{Host-Galaxy}, we classified the host galaxy of PS1-13cbe to be a Seyfert using the SDSS spectrum. Moreover, the host galaxy can be classified as a Seyfert 1.9/2 galaxy since it just shows narrow emission lines and possible weak broad H$\alpha$ without any sign of a broad H$\beta$ line, as shown in the left panels of Figures~\ref{Halpha} and ~\ref{Hbeta}. At the time of the outburst, the spectrum of PS1-13cbe taken with LDSS3 showed broad H$\alpha$ and H$\beta$ components with FWHM velocities of $3385 \pm 32$ km~s$^{-1}$ and $3277 \pm 110$ km~s$^{-1}$, respectively. These high values of FWHM are the sign of the high velocity, dense, and highly ionised gas clouds in the Broad Line Region (BLR) close to the central black hole (BH) where the broad emission lines originate from. The presence of these broad lines suggests that the galaxy transformed from a Seyfert 1.9/2 to a Seyfert 1 galaxy. On the other hand, we did not detect broad emission lines in the spectra taken with OSMOS and that means the host galaxy of PS1-13cbe transformed back from a Seyfert 1 to a Seyfert 2 galaxy in less than 2 years and continues in that state as there are no signs of broad lines in spectra taken with OSMOS $\sim 4$ years after the peak of the outburst. 

\begin{table}
    \renewcommand\thetable{2}
	\captionsetup{justification=centering}
	\centering \caption{Luminosity of Broad Lines}
	\begin{tabular}{l ccccr}
		\toprule
		 Date  (UT) & Instrument & Broad H$\alpha$ & Broad H$\beta$ \\
		\midrule
		2003 June 26 & SDSS  & 4.87 $\pm$ 0.37 & 1.25 $\pm$ 0.39  \\	
		2013 Oct  05 & LDSS3  & 16 $\pm$ 0.28 & 4.39 $\pm$ 0.22  \\
		2015 Oct  03 & OSMOS & 3.92 $\pm$ 0.45 & < 1.2      \\
		2016 Nov 16& OSMOS &    & < 0.85       \\
		2017 June 18& OSMOS &  < 0.86   &  < 1.14       \\
		\bottomrule
		\
	\end{tabular}

	\raggedright \textbf{Note:} Luminosity of broad H$\alpha$ and H$\beta$ lines in SDSS, LDSS3, and OSMOS spectra reported in units of $10^{40}$ erg~s$^{-1}$. Non-detections  are reported as $3\sigma$ upper limits.
	\label{table:2}
\end{table}

We estimate the mass of the central BH using two methods. The first method uses the stellar kinematics and the revised scaling relation between the SMBH mass and stellar velocity dispersion \mbox{\citep{mcconnell2013revisiting}}, where $\sigma =  93.52$ km~s$^{-1}$ was provided from the SDSS spectrum of the host galaxy, and results in an estimate of M$_{\mathrm{BH}}= 2.9^{+3.5}_{-1.7} \times 10^{6}$ M\textsubscript{$\odot$}. However, this estimation is subject to uncertainties because of the large scatter and lack of constraints at the low $\sigma$ and M$_{\mathrm{BH}}$ region of the M$_{\mathrm{BH}}-\sigma$ scaling relation, particularly in a case where we do not have a decomposition of the galaxy that separates the bulge component \citep{mcconnell2013revisiting, kormendy2013coevolution}. 
%\textcolor{red}{large scatter and lack of constraints at low-sigma end of diagram, either mcconnell \& ma or the ho\&kormendy annual reviews article:
%https://ui.adsabs.harvard.edu/\#abs/2013ARA\&A..51..511K/abstract
%}

The mass of the SMBH can also be estimated using photoionisation equilibrium, by applying a mass-scaling relationship based on the FWHM of the broad H$\beta$ emission line and continuum luminosity. Therefore, using the measured FWHM $\approx 3277$ km~s$^{-1}$, intrinsic luminosity at $5100 \angstrom$ ($\lambda L_{5100}$) of $(6.2 \pm 0.2)\times 10^{42}$ erg~s$^{-1}$ at the time of the spectrum, and the mass-scaling relationship of \cite{vestergaard2006determining}, we calculate the mass of the central BH to be $(2.2\pm 0.1) \times 10^{7}$ M\textsubscript{$\odot$}. We note that the values used in this calculation are measured during the outburst and can be affected by the changes in the accretion structure and also rely on the assumption that the (unknown) shape of the ionising continuum is similar to those objects used to calibrate the scaling relationship. We prefer this mass estimate in our calculations below because we believe the assumptions behind this photoionisation calculation to be more robust.

The Eddington luminosity for this BH mass is $L_{\mathrm{Edd}} = (2.7 \pm 0.1) \times 10^{45}$ erg~s$^{-1}$. We also calculate the intrinsic luminosity at $5100 \angstrom$ ($\lambda L_{5100}$) for PS1-13cbe that is $(1.16 \pm 0.01)\times 10^{43}$ erg~s$^{-1}$ at the time of the peak and then we estimate the bolometric luminosity using $\lambda L_{\lambda}$ and a conversion factor of 8.1 \citep{runnoe2012updating} to convert from monochromatic to bolometric luminosity to be $(9.4 \pm 0.1) \times 10^{43}$ erg~s$^{-1}$. This results in an Eddington parameter of $\lambda_{\mathrm{Edd}} = L_{\mathrm{bol}}/L_{\mathrm{Edd}} \approx 0.03$ at the peak of the outburst.

\section{Interpretation of the features of PS1-13cbe}\label{source}
\begin{figure*}
	\includegraphics[scale = 0.80]{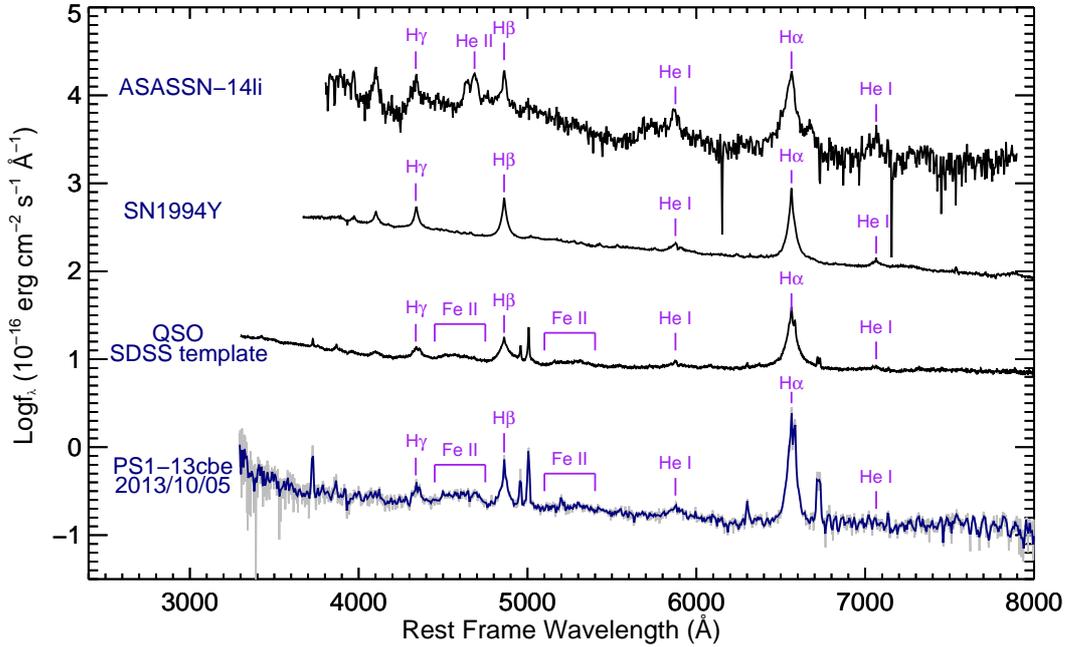}
	\caption{Optical spectrum of PS1-13cbe taken with LDSS3 during the outburst after subtracting the stellar continuum (blue). The spectra of the comparison objects include a QSO \citep[SDSS QSO;][]{berk2001composite2} template, a SN Type IIn \citep[SN1994Y;][]{filippenko1997optical} and a TDE \citep[ASASSN-14li;][]{holoien2015six} from bottom to top (black), the emission lines are labeled (purple).}
	\label{Comparison}
\end{figure*}

In Figure~\ref{Comparison}, we show the spectrum of PS1-13cbe at the time of the outburst (after subtraction of the star light component from the host galaxy model) alongside the comparison objects, including a QSO (SDSS QSO template), a Type IIn SN \citep[SN1994Y;][]{filippenko1997optical}, and a tidal disruption event (TDE: ASASSN-14li; \citealt{holoien2015six}). All of these objects show the presence of broad H$\alpha$ and H$\beta$ lines with emission line profiles similar to the PS1-13cbe spectrum. It is notable that the PS1-13cbe spectrum closely resembles that of the QSO template, which is consistent with AGN activity. However, we estimate the spectral index at the epoch of the LDSS3 spectrum to be $\alpha = - 0.58$, which is redder than the highly variable QSOs studied in \cite{wilhite2005spectral} that had spectral index of $ \alpha =-2$ for the average difference spectrum (bright phase minus faint phase). 

In this section, we discuss SNe, TDE, and AGN variability as three possible interpretations for PS1-13cbe. However, before discussing the details about these scenarios we have summarized the key features of PS1-13cbe: 
\begin{itemize}
\item PS1-13cbe occurred in the nucleus of a Seyfert 2 galaxy with a central SMBH with mass of $\sim 2 \times 10^{7}$ M\textsubscript{$\odot$}. 
\item PS1-13cbe brightened in the course of $\sim 70$ days and reached a peak total optical luminosity of $(1.06 \pm 0.01) \times 10^{43}$ erg~s$^{-1}$. 
\item The temperature of PS1-13cbe roughly stayed constant and $g-r$ (transient component) did not show any colour evolution.
\item The spectra of PS1-13cbe show significant evolution over the course of  $\sim 12$ years where broad H$\alpha$, H$\beta$, and H$\gamma$ lines appear and disappear from the spectra. 
\end{itemize}

\subsection{Type IIn Supernovae interpretation of PS1-13cbe} 

Type IIn SNe show broad H Balmer lines and blue continua in their optical spectra that can resemble the spectra of AGNs in specific phases of their evolution \citep{filippenko1989seyfert}. As we show in Figure~\ref{Comparison}, the lack of P Cygni profiles in the spectrum of PS1-13cbe is similar to the spectral features of Type IIn SNe such as SN1994Y \citep{filippenko1997optical} that is used as an example here for the purpose of the comparison. Additionally, the peak luminosity of PS1-13cbe, $M_{r} \approx -19.4$, is not substantially brighter than the typical Type IIn SNe, which have a broad peak luminosity range of $-18.5 \le M_{r} \le -17$ (e.g., \citealt{kiewe2011caltech}).

The Type-IIn SN hypothesis is however disfavored for the following reasons. First, the narrow emission line ratios in Figure~\ref{Host_LineRatio} clearly identify the galaxy as a Seyfert, which increases the chances that this variability was caused by the existing AGN rather than a SN. Furthermore, the temperature of PS1-13cbe is roughly constant and it does not show any colour evolution, which is not consistent with the cooling of the ejecta typically seen in SNe. Additionally, the double hump behavior seen in the light curves (Figure~\ref{PS1_lightCurve} and Figure~\ref{LuminosityVsTime}) of PS1-13cbe is not commonly observed in Type IIn SNe \citep[e.g.,][]{kiewe2011caltech, taddia2013carnegie}. While Type II SNe have been very rarely observed to exhibit multiple rebrightening bumps after the first peak \citep[e.g., iPTF14hls and iPTF13z;][]{arcavi17,nyholm2017bumpy}, the small fluctuations observed in the light curve of PS1-13cbe are not present in their light curves and the rise and rebrightening timescales in the case of PS1-13cbe are much shorter in comparison.  

Based on the reasons provided in this section, we disfavor the Type IIn SN origin of PS1-13cbe, but this possibility cannot be completely ruled out.  

\subsection{PS1-13cbe as a TDE}\label{TDE_Poss}
Most of the optically detected TDEs show a lack of colour evolution and constant blackbody temperature, consistent with our observations of PS1-13cbe \citep{gezari2012ultraviolet, chornock2013ultraviolet}. Furthermore, broad emission lines such as H$\alpha$, H$\beta$ and He II have been detected in the spectra of the TDEs and disappeared at later times \citep{arcavi2014continuum, van2011optical}. Additionally, broad H$\alpha$ and H$\beta$ lines are detected in the spectrum close to the time of the peak and either were not present (H$\beta$) or only weakly present (H$\alpha$) in earlier spectra (SDSS) and disappear in the later spectra (OSMOS). Therefore, it is plausible that PS1-13cbe has a TDE origin. However, we do not have optical photometry data on the decline to check whether the light-curve decay at the predicted rate of $t^{-5/3}$ for TDEs \citep{rees1988tidal, evans1989tidal}. The lack of any X-ray observations at the time of the outburst is another limitation we face to investigate the TDE and SN scenarios. 

As seen in Figure~\ref{Host_LineRatio}, the host of PS1-13cbe is classified as a Seyfert galaxy, which points to the presence of a pre-existing AGN. However, TDEs can happen in AGN galaxies, and in fact it has been suggested that TDEs may prefer galaxies with pre-existing steady accretion to their central BH \citep{hills1975possible, blanchard2017ps16dtm} where the dense star-formation clouds and the pre-existing accretion disk can increase the chance of the tidal encounter with the stars and reduce the relaxation time \citep{perets2006massive, blanchard2017ps16dtm}. For example, the spectrum of the hosts of the optical/UV TDE SDSS J0748 \citep{wang2011transient} and optical TDE candidate PS16dtm \citep{blanchard2017ps16dtm} show weak and narrow emission lines that might be the result of AGN presence in the core. Similar to them, the host of the TDE ASASSN-14li \citep{holoien2015six} also shows traces of  ongoing weak AGN activity at the center \citep{van_Velzen62, alexander2016discovery}.

However, we disfavor the TDE interpretation of PS1-13cbe because of the following reasons. The blackbody temperatures in optically-selected TDEs typically range from $\sim20000$ K \citep[e.g., PS1-11af;][]{chornock2013ultraviolet} to $\sim35000$ K and higher \citep[e.g., ASASSN-14li;][]{holoien2015six}, while the continuum of PS1-13cbe at the time of the outburst is not as blue as a typical TDE and, as we show in Figure~\ref{T_BB}, the inferred blackbody temperature of PS1-13cbe is no more than $\sim 9000$ K. Moreover, the He II $\lambda4686$ line that is frequently seen in the spectra of TDEs (e.g., ASASSN-14li in Figure \ref{Comparison}), is not visible in the spectrum of  PS1-13cbe that was taken close to the time of outburst. However, it is worth mentioning that there are TDE candidates that lack the presence of the He II $\lambda4686$. Additionally, the rebrightening of PS1-13cbe by almost 75\% is not typically seen in TDEs.  However, it is notable that the TDE candidate ASASSN-15lh showed re-brightening in the UV +60 days after the peak \citep{dong2016asassn, leloudas2016superluminous, margutti2017x}. The optical TDE candidate PS16dtm also showed dimming and rebrightening $\sim 150$ days after the beginning of the rise \citep{blanchard2017ps16dtm}. 

Furthermore, the light curves of the PS1-13cbe have noticeable fluctuations on few days timescales (such as the one visible in $gri_{\mathrm{P1}}$ near $-20$ days in Figure~\ref{PS1_lightCurve}) that are inconsistent with the observed smooth optical light curves of most of the other known optically-selected TDEs \citep[e.g.,][]{van2011optical, gezari2012ultraviolet, chornock2013ultraviolet, arcavi2014continuum, holoien2014asassn}. However, noticeable fluctuations have been observed in the light curves of optical TDE candidate PS16dtm \citep{blanchard2017ps16dtm}.

The optical and spectral features of PS1-13cbe and the reasons provided in this section show that the TDE origin of  PS1-13cbe is a possible but not a likely scenario. 

\subsection{PS1-13cbe as a ``Changing Look" AGN}
Recently, a new type of AGN variability was discovered in objects called ``changing look" AGNs that show the appearance or disappearance of broad emission lines followed by an order of magnitude increase or decrease in the continuum  and change type from Type 1 to Type 1.8, 1.9 or vice versa \citep[e.g., ][]{runnoe2016now, shappee2014man, lamassa2015discovery, gezari2016iptf, macleod2016systematic}. This CL behavior observed in some AGNs can be caused by at least three mechanisms. In the first scenario, variation of the obscuration when material such as dust clouds outside of the BLR move in or out of the line of sight that can obscure or clear the view to the BLR \citep{elitzur2012unification}. Another mechanism can be variations of accretion rate that transforms the structure of the BLR. The AGN will transfer from Type 1 when the accretion rate is high and the broad lines are visible to a Type 2 when the accretion rate is low and the broad lines cannot exist or vice versa \citep{elitzur2014evolution}. Furthermore, it has been suggested that transient events such as TDEs can cause this CL behavior \citep{eracleous1995elliptical}.

As shown in Figure~\ref{Halpha} and ~\ref{Hbeta}, we only detected the presence of weak broad H$\alpha$ and no broad H$\beta$ emission line in the spectrum taken by SDSS. However, the strong broad H$\alpha$ line and H$\beta$ lines appeared at the time of the outburst that followed an observed increase in the flux and disappeared again in the spectra taken at later times. This shows that the AGN changed type from a Type 1.9/2 to a Type 1 and then to a Type 2 because no broad emission lines were detected in later observations. 

We estimate the luminosity at $5100 \, \angstrom$ ($\lambda L_{5100}$) based on our estimate that the non-stellar continuum contributes $\le 10 \%$ in the SDSS spectrum to be $ \le 0.15 \times 10^{43}$ erg~s$^{-1}$ in quiescence. Additionally, we measured the intrinsic $5100 \, \angstrom$ ($\lambda L_{5100}$) luminosity at the peak of the outburst to be $1.16\times 10^{43}$ erg~s$^{-1}$, which shows a factor $\gtrsim 8$ increase in optical luminosity. Also, during this time the broad H$\alpha$ emission varied by a factor of $\sim 4$ in flux.

The observed luminosity changes and appearance and disappearance of the H$\alpha$ and H$\beta$ lines can be caused by one of the three mentioned mechanisms that are discussed in more detail in the following sections.  

\subsubsection{Obscuration of the AGN}
The CL behavior seen in PS1-13cbe can be caused by intervening material which is located outside of the BLR and orbits on a Keplerian orbit that can obscure or give a clear view to the BLR by moving in or out of the line of sight. 
%The higher estimated bolometric luminosities of $\sim 1.6 \times 10^{44}$ erg~s$^{-1}$ at the time of quiescence compared to the bolometric luminosity of $\sim 9.4 \times 10^{43}$ erg~s$^{-1}$ estimated at the time of outburst and 
The bolometric luminosity of $\sim 9.4 \times 10^{43}$ erg~s$^{-1}$ estimated at the time of outburst using bolometric corrections for unobscured AGNs is less than the bolometric luminosity of $\sim 1.6 \times 10^{44}$ erg~s$^{-1}$ at the time of quiescence estimated from the narrow $\lambda$5007 and mid-IR excess, which possibly indicates that the AGN is not fully unobscured at the peak of the outburst.
In addition, the redder spectrum compared to the QSO template suggests such a scenario to be possible. However, we disfavor the changing obscuration scenario in the case of PS1-13cbe for the following reasons. First, we estimate the characteristic radius of the BLR using the \textit{R-L} relation calibrated presented by \citep{bentz2013low} to be $R_{BLR} \sim 11$ light days. Then, using the relation for crossing time presented by \cite{lamassa2015discovery}, we estimate the crossing time for an obscuring object orbiting on a Keplerian orbit outside of BLR. In the most ideal case that minimizes the crossing time, the obscuring object should be at $r_{orb} \ge R_{BLR} = 11$ light days so that it can intercept a substantial amount of the broad Balmer flux from the BLR. Even in the extreme case of assuming $r_{orb} = R_{BLR}$, the crossing time for the obscuring object is $\sim 23$ years. In a more realistic scenario where $r_{orb} \ge 3R_{BLR}$ the crossing time is even higher. Not only are these timescales too long to explain the CL behavior of  PS1-13cbe, but also the existence of intervening material with physical properties that can obscure the whole region of the continuum and the BLR at such radii is not obvious. 

One of the possible candidates for obscuration is the torus that lies just outside of the BLR beyond the dust sublimation radius \citep[e.g., ][]{suganuma2006reverberation, koshida2014reverberation, lamassa2015discovery}. However, the sublimation radius is itself 4-5 times larger than the $R_{BLR}$ where the crossing timescale is $\sim 29$ years, which is again too long to explain the observed event. Another possible obscuring scenario is that the dimming of the continuum itself promotes the formation of the dust that is able to cover the BLR. However, the timescale for such dust formation with the gas density of $\sim 10^{5}$ cm$^{-3}$ in the torus \citep[e.g.,][]{nenkova2002dust} is $\sim 10^{3}$ years \citep{draine2009interstellar, lamassa2015discovery} that is far too long for this scenario to be true in the case of PS1-13cbe. 

In addition, if the dust obscuration scenario is true, then the change in luminosity should follow the colour evolution and thus the colour evolution and luminosity change should be correlated. By contrast, there is no colour evolution while the luminosity is changing in the case of PS1-13cbe. We fit a linear regression model to luminosity versus $g-r$ colour index and found a slope of $-0.01 \pm 0.01$ with a coefficient of determination $R^{2} \sim 0.056$. These results confirm the lack of correlation between luminosity and $g-r$ colour index.

\subsubsection{Tidal Disruption Events}
As we mentioned in the the Section~\ref{TDE_Poss}, the lack of the colour evolution, constant blackbody temperature, and appearance/disappearance of the broad H$\alpha$ and H$\beta$ lines suggest that the optical outburst and the appearance of the apparent blue continuum may have been powered by a TDE in the pre-existing AGN. Based on the reasons that we provided in the Section~\ref{TDE_Poss}, we conclude that TDE origin of the outburst in PS1-13cbe is not favored, but it cannot be ruled out completely. 

\subsubsection{Accretion disk instabilities}
In the light curves of  PS1-13cbe shown in Figure~\ref{PS1_lightCurve}, there are small undulations that can be seen in all of the filters (e.g., near $-20$ days). The amplitudes of these changes are consistent with the variability observed in AGNs \citep[e.g.,][]{macleod2012description} and point to fluctuations in ongoing accretion activity. 

It has been shown that in AGNs the optical/UV emission is generated in the accretion disk, with two possible classes of models for the propagation of fluctuations. One is that the locally generated viscous perturbations can produce local blackbody emission \mbox{\citep[e.g., ][]{krolik1999book, liu2008tests}}. In the case of these so-called ``outside-in" variations that are produced by changes in accretion rate, the accretion flow fluctuations propagating inward and across the accretion disk first affect the optical region located at outer radii and then affect the UV and X-ray emitting regions. Another origin can be the reprocessing of the UV or X-rays \citep[e.g.,][]{krolik1991ultraviolet, cackett2007testing, liu2008tests}. In this case, the X-rays from the central source irradiate the disk and produce ``inside-out" variations from short to long wavelengths \citep{lamassa2015discovery, shappee2013swift}. 

The timescale over which the accretion changes happen that might produce ``outside-in" variations is known as the inflow timescale. More accurately, changes in the accretion responsible for the ``outside-in" variations happen on the inflow time scale, which is the time it takes a parcel of gas in a given radius in the accretion disk to radially move to the center. Assuming the optical continuum emission radius of $R \approx 200\, r_{S}$ \citep[e.g.,][]{morgan2010quasar, fausnaugh2016space} and using the relationship between radius and inflow timescale presented by \cite{lamassa2015discovery}, we calculate the inflow timescale to be $t_{\rm infl} \sim 10^{6}$ years, which is much longer than the observed change in the continuum flux of PS1-13cbe. 

However, it is notable that the optical continuum originates in a part of the disk where the radius is an order of magnitude larger than the UV-emitting region, which results in a several thousand times larger inflow timescale than for the UV-emitting region \citep{lamassa2015discovery}. Thus, we conclude that the rapid continuum flux change in PS1-13cbe is too short to be generated by outside-in variations (perturbations in a given radius of the disk that propagate radially inward), but might be more characteristic of a disturbance in the inner accretion flow that propagates outward. 

Assuming a standard thin disk model \citep{shakura1973black} with an optically thick and geometrically thin accretion disk, we calculate the dynamical (orbital) time-scale $t_{dyn} =  1/\Omega$ of the SMBH where  $\Omega = \sqrt{\frac{GM}{R^{3}}}$. We can rewrite the dynamical time-scale to be $t_{dyn} = 2^{3/2}\frac{GM}{c^3} \Big( \frac{R}{r_{S}}\Big)^{3/2}$ where $c$ is the speed of light and $r_{S} = \frac{2GM}{c^2}$ is the Schwarzschild radius of the central BH. The dynamical timescale around a SMBH with mass of $2.2 \times 10^{7}$~M$_{\odot}$ can be written as $t_{dyn} \approx 310 \Big( \frac{R}{r_{S}}\Big)^{3/2}$ s. Assuming an optical emission distance of $R \approx 200\,r_{S}$ \citep[e.g.,][]{morgan2010quasar, fausnaugh2016space}, we calculate the dynamical timescales of $\approx 10$ days. Then, using using $t_{th} = \alpha^{-1}t_{dyn}$ we calculate thermal timescale to be $\approx 99$ days. We note that these values are not strongly dependent on the uncertainties in the BH mass for this object, so the ordering of timescales is robust.

Among all of the calculated timescales, only the thermal timescale of $\sim 99$ days is reasonably similar to the observed timescale of $\sim 70$ days in the case of PS1-13cbe and suggests another possible scenario where reprocessing of the UV or X-rays can produce the optical variations. In this case, an increase in the X-rays that originate in the smaller hot corona that is closer to the central BH can heat the inner part of the accretion disk first, move outward, and generate inside-out variations by irradiating the disk and driving an increase in the blue and then red emission. This scenario has been observed and well studied in NGC 2617 \citep{shappee2013swift}. \cite{shappee2013swift} detected the variability first in X-rays and then with time lags in UV and NIR and concluded that the observed continuum flux variability resulted from inside-out variations. Also, NGC 2617 is the only case where it was clearly observed that X-ray variability drove UV-NIR variability.  

\subsection{Comparison to other Changing Look AGNs}
PS1-13cbe is one of the few CL AGNs that has been observed during the turn-on phase \citep{cohen1986variability, storchi1993double, aretxaga1999seyfert, eracleous2001ngc, shappee2014man, gezari2016iptf} by suddenly demonstrating the appearance of broad H$\alpha$ and H$\beta$ emission lines. The CL behavior has been observed in other candidates where the broad lines appeared or disappeared  in spectra taken a couple of years to decades apart \citep[e.g.,][]{lamassa2015discovery, runnoe2016now,macleod2018}. By contrast, in PS1-13cbe the outburst timescale is very short. PS1-13cbe ``turned-on" in only $70$ days and the broad lines were observed in a spectrum taken $57$ days after the peak of the outburst. Furthermore, the broad lines disappeared before two years after the time of the peak which is again very short and never re-appeared in later spectra (see Figure~\ref{PS1_spectra}). \cite{gezari2016iptf} also presented the CL quasar iPTF 16bco that had a ``turn-on" timescale of $\leq 1$ year which is very short for a BH with $M_{BH} \sim 10^{8}\,M_{\odot}$ compared to others. PS1-13cbe and iPTF 16bco are the only CLs that demonstrate extremely short turn-on timescales which push the limit of the accretion disk theory. One more interesting fact about PS1-13cbe is that the light curves showed a double peak behavior observed in all of the optical bands (see Figure \ref{PS1_lightCurve}). This behavior was also observed in the X-rays/UV in the case of NGC 2617 \citep{shappee2014man}, in the UV/optical in the case of the ASASSN-15lh TDE candidate \citep{dong2016asassn, leloudas2016superluminous, margutti2017x}, and in the optical band in the case of the PS16dtm TDE candidate \citep{blanchard2017ps16dtm}.  

\section{Conclusions}\label{Conclusion}
We present a transient event that was discovered in the PS1/MDS survey, PS1-13cbe, at redshift $z = 0.12355$. The outburst happened in the nucleus of a galaxy that is classified as a Seyfert 2 (see Figure~\ref{Host_LineRatio}) using the SDSS archival data that was taken a decade before the outburst. At the time of the outburst, the galaxy changed type to a Seyfert 1 as broad H$\alpha$ and H$\beta$ appeared and the continuum brightened in the spectrum taken with LDSS3 $+57$ days after the peak and then changed its type back to a Seyfert 2 as the broad H$\alpha$ and H$\beta$ disappeared in the spectrum taken with OSMOS 2 years later and did not reappear in spectra taken 3 and 4 years after the outburst. The optical photometry shows that the continuum flux increased by a factor of $\sim$8 on a timescale of $\sim 70$ days and declined for next $\sim 50$ days and then rose again over the course of the next $\sim 50$ days. 

Observational evidence presented in this work argues against the Type IIn SN and TDE interpretations. The constant colour evolution and blackbody temperature during the outburst and also the presence of a pre-existing AGN disfavour the SN Type IIn scenario. As mentioned, TDEs have been observed in the galaxies with pre-existing AGNs \citep[e.g., SDSS J0748; ][]{wang2011transient}; however, the lack of a broad He II$\lambda4686$ emission line, low blackbody temperature at the time of peak, a light curve that has small fluctuations, and unusual re-brightening by 75\% are inconsistent with properties of known TDEs.

We conclude that PS1-13cbe is a changing-look AGN that has been powered by instabilities in the accretion disk. We argued against the obscuration scenario and TDE origin of these accretion disk instabilities by showing that the crossing and viscous timescales are longer than the timescale observed in the case of PS1-13cbe. Furthermore, we also argued against outside-in variations by calculating the inflow timescale which is too long to explain the observed timescale here. 

We also conclude that the thermal instabilities in the accretion disk are most likely the source of the outburst and CL behavior. These thermal instabilities may have caused inside-out variations that have generated the observed optical variability in the light curves of PS1-13cbe. One very interesting point about PS1-13cbe is that the observed turn-on timescale pushes the limits of viscous accretion disk theory which predicts much longer timescales and it might be one of the CLs that have shown the most rapid change of the state \citep[iPTF 16bco; ][]{gezari2017iptf} compared to the other CLs. 

Other CLs have been observed over a timespan of years and sometimes decades apart; however, the short timescale observed here suggests that other candidates may have been through these short timescale outbursts. Therefore, more frequent observations with a higher cadence and multiwavelength coverage to overcome limitations such as the one we faced here with the lack of X-ray observations at the time of outburst, will provide us more insight to better understand the changing look behavior of AGNs.

\section*{Acknowledgements}

%The Acknowledgements section is not numbered. Here you can thank helpful
%colleagues, acknowledge funding agencies, telescopes and facilities used etc.
%Try to keep it short.

%Standard acknowledgments for: PS1, MDM, Magellan, SDSS.

The Pan-STARRS1 Surveys (PS1) and the PS1 public science archive have been made possible through contributions by the Institute for Astronomy, the University of Hawaii, the Pan-STARRS Project Office, the Max-Planck Society and its participating institutes, the Max Planck Institute for Astronomy, Heidelberg and the Max Planck Institute for Extraterrestrial Physics, Garching, The Johns Hopkins University, Durham University, the University of Edinburgh, the Queen's University Belfast, the Harvard-Smithsonian Center for Astrophysics, the Las Cumbres Observatory Global Telescope Network Incorporated, the National Central University of Taiwan, the Space Telescope Science Institute, the National Aeronautics and Space Administration under Grant No. NNX08AR22G issued through the Planetary Science Division of the NASA Science Mission Directorate, the National Science Foundation Grant No. AST-1238877, the University of Maryland, Eotvos Lorand University (ELTE), the Los Alamos National Laboratory, and the Gordon and Betty Moore Foundation.

We thank S.B. Cenko and the {\it Swift} team for approving and executing our {\it Swift} Target of Opportunity observations.
This work is based in part on observations obtained at the MDM Observatory, operated by Dartmouth College, Columbia University, Ohio State University, Ohio University, and the University of Michigan.
This paper includes data gathered with the 6.5 meter Magellan Telescopes located at Las Campanas Observatory, Chile.

Some of the computations presented in this work were performed on Harvard University's Odyssey computer cluster, which is maintained by the Research Computing Group within the Faculty of Arts and Sciences.

Funding for the SDSS and SDSS-II has been provided by the Alfred P. Sloan Foundation, the Participating Institutions, the National Science Foundation, the U.S. Department of Energy, the National Aeronautics and Space Administration, the Japanese Monbukagakusho, the Max Planck Society, and the Higher Education Funding Council for England. The SDSS website is \url{http://www.sdss.org/}. The SDSS is managed by the Astrophysical Research Consortium for the Participating Institutions. The Participating Institutions are the American Museum of Natural History, Astrophysical Institute Potsdam, University of Basel, University of Cambridge, Case Western Reserve University, University of Chicago, Drexel University, Fermilab, the Institute for Advanced Study, the Japan Participation Group, Johns Hopkins University, the Joint Institute for Nuclear Astrophysics, the Kavli Institute for Particle Astrophysics and Cosmology, the Korean Scientist Group, the Chinese Academy of Sciences (LAMOST), Los Alamos National Laboratory, the Max-Planck-Institute for Astronomy (MPIA), the Max-Planck-Institute for Astrophysics (MPA), New Mexico State University, Ohio State University, University of Pittsburgh, University of Portsmouth, Princeton University, the United States Naval Observatory, and the University of Washington.

%%%%%%%%%%%%%%%%%%%%%%%%%%%%%%%%%%%%%%%%%%%%%%%%%%

%%%%%%%%%%%%%%%%%%%% REFERENCES %%%%%%%%%%%%%%%%%%

% The best way to enter references is to use BibTeX:
\bibliographystyle{mnras.bst}
\bibliography{bibfile} % if your bibtex file is called example.bib
\begin{table}
    \renewcommand\thetable{1}
	\captionsetup{justification=centering}
	\centering \caption{Photometry of PS1-13cbe.}
	\begin{tabular}{clcccc}
		\toprule
        Data of Observation & \multicolumn{1}{l}{Epoch} & Filter & Mag  & Mag \\
        (MJD) & \multicolumn{1}{c}{(days)} & & (observed) &  Uncertainty\\
        \midrule
		56217.3 & -295.3 & $g_{p1}$ &(24.32) & \\
		56475.5 & -37.1	 & $g_{p1}$ & 19.89 & 0.03 \\
		56480.5 & -32.1	 & $g_{p1}$ & 19.79 & 0.03 \\
		56486.6 & -26.0	 & $g_{p1}$ & 19.76 & 0.03 \\
		56489.5 & -23.1	 & $g_{p1}$ & 19.91 & 0.03 \\
		56501.4 & -11.2	 & $g_{p1}$ & 19.79 & 0.03 \\
		56507.6 & -5.0	 & $g_{p1}$ & 19.76 & 0.03 \\
		56512.5 & -0.1   & $g_{p1}$ & 19.66 & 0.03 \\
		56518.5 & 5.9    & $g_{p1}$ & 19.85 & 0.03 \\
		56531.5 & 18.9   & $g_{p1}$ & 20.16 & 0.04 \\
		56534.5 & 21.9   & $g_{p1}$ & 20.20 & 0.04 \\
		56539.4 & 26.8   & $g_{p1}$ & 20.30 & 0.04 \\
		56545.3 & 32.7   & $g_{p1}$ & 20.38 & 0.04 \\
		56548.3 & 35.7   & $g_{p1}$ & 20.42 & 0.06 \\
		56563.3 & 50.7   & $g_{p1}$ & 20.49 & 0.05 \\
		56571.2 & 58.6   & $g_{p1}$ & 20.50 & 0.04 \\
		56574.2 & 61.6   & $g_{p1}$ & 20.43 & 0.06 \\
		56590.3 & 77.7   & $g_{p1}$ & 20.35 & 0.04 \\
		56593.3 & 80.7   & $g_{p1}$ & 20.24 & 0.04 \\
		56596.2 & 83.6   & $g_{p1}$ & 20.22 & 0.04 \\
		56601.2 & 88.6   & $g_{p1}$ & 20.34 & 0.07 \\
		56628.3 & 115.7  & $g_{p1}$ & 20.11& 0.17 \\
		56285.2 & -227.4& $r_{p1}$ &(24.32)&         \\
		56475.6 & -37.0	& $r_{p1}$ & 19.85 & 0.04 \\
		56479.5 & -33.1	& $r_{p1}$ & 19.75 & 0.04 \\
		56486.6 & -26.0	& $r_{p1}$ & 19.66 & 0.03 \\
		56489.5 & -23.1	& $r_{p1}$ & 19.80 & 0.03 \\
		56501.4 & -11.2	& $r_{p1}$ & 19.70 & 0.03 \\
		56507.6 & -5.0	& $r_{p1}$ & 19.64 & 0.03 \\
		56508.5 & -4.1	& $r_{p1}$ & 19.63 & 0.03 \\
		56512.6 & 0.0	& $r_{p1}$ & 19.60 & 0.03 \\
		56518.5 & 5.9   & $r_{p1}$ & 19.72 & 0.03 \\
		56531.5 & 18.9	& $r_{p1}$ & 20.04 & 0.04 \\
		56534.5 & 21.9	& $r_{p1}$ & 20.08 & 0.04 \\
		56538.3 & 25.7	& $r_{p1}$ & 20.03 & 0.05 \\
		56545.3 & 32.7	& $r_{p1}$ & 20.11 & 0.04 \\
		56571.2 & 58.6	& $r_{p1}$ & 20.44 & 0.06 \\
		56574.3 & 61.7	& $r_{p1}$ & 19.81 & 0.37 \\
		56590.3 & 77.7	& $r_{p1}$ & 20.29 & 0.05 \\
		56593.3 & 80.7	& $r_{p1}$ & 20.11 & 0.05 \\
		56596.3 & 83.7	& $r_{p1}$ & 20.05 & 0.04 \\
		56597.3 & 84.7	& $r_{p1}$ & 20.08 & 0.04 \\
		56626.2 & 113.6	& $r_{p1}$ & 20.11 & 0.07 \\
		56284.2 & -228.4 & $i_{p1}$ & (22.65) &      \\
		56473.5 & -39.1	 & $i_{p1}$ & 19.65 & 0.04 \\
		56476.6 & -36.0	 & $i_{p1}$ & 19.60 & 0.04 \\
		56481.6 & -31.0	 & $i_{p1}$ & 19.51 & 0.05 \\
		56484.5 & -28.1	 & $i_{p1}$ & 19.42 & 0.03 \\
		56491.5 & -21.1	 & $i_{p1}$ & 19.54 & 0.03 \\
		56505.4 & -7.2	 & $i_{p1}$ & 19.46 & 0.03 \\
		56508.6 & -4.0	 & $i_{p1}$ & 19.44 & 0.03 \\
		56513.6 & 1.0	 & $i_{p1}$ & 19.44 & 0.03 \\
		56516.5 & 3.9	 & $i_{p1}$ & 19.46 & 0.03 \\
		56520.5 & 7.9	 & $i_{p1}$ & 19.54 & 0.03 \\
		56532.4 & 19.8	 & $i_{p1}$ & 19.79 & 0.04 \\
		56538.4 & 25.8	 & $i_{p1}$ & 19.72 & 0.04 \\
		56540.4 & 27.8	 & $i_{p1}$ & 19.70 & 0.04 \\
		56550.5 & 37.9	 & $i_{p1}$ & 20.18 & 0.06 \\
		56558.4 & 45.8	 & $i_{p1}$ & 20.07 & 0.05 \\
		56564.4 & 51.8	 & $i_{p1}$ & 19.96 & 0.05 \\
		56567.2 & 54.6	 & $i_{p1}$ & 20.10 & 0.06 \\
		56569.3 & 56.7	 & $i_{p1}$ & 20.02 & 0.05 \\
		\bottomrule
	\end{tabular}
	\label{phot_tab}
\end{table}

\begin{table}
%\captionsetup{justification=centering}
%\contcaption{}
\renewcommand\thetable{1}
\centering \caption{\textit{continued}}
\label{tab:continued}
\begin{tabular}{clcccc}
		\toprule
        Data of Observation & \multicolumn{1}{l}{Epoch} & Filter & Mag  & Mag \\
        (MJD) & \multicolumn{1}{c}{(days)} & & (observed) &  Uncertainty\\
        \midrule
        56575.3 & 62.7	 & $i_{p1}$ & 20.01 & 0.05 \\
        56588.4 & 75.8	 & $i_{p1}$ & 19.79 & 0.04 \\
		56597.2 & 84.6	 & $i_{p1}$ & 19.88 & 0.04 \\
		56599.3 & 86.7	 & $i_{p1}$ & 19.89 & 0.04 \\
		56626.2 & 113.6	 & $i_{p1}$ & 19.85 & 0.06 \\
		56216.2 & -296.4 & $z_{p1}$ & (22.91) &      \\
		56453.5 & -59.1	 & $z_{p1}$ & 19.86 & 0.05 \\
		56456.6 & -56.0	 & $z_{p1}$ & 19.71 & 0.04 \\
		56474.6 & -38.0	 & $z_{p1}$ & 19.58 & 0.05 \\
		56477.5 & -35.1	 & $z_{p1}$ & 19.67 & 0.04 \\
		56480.5 & -32.1	 & $z_{p1}$ & 19.44 & 0.04 \\
		56482.5 & -30.1	 & $z_{p1}$ & 19.41 & 0.03 \\
		56485.6 & -27.0	 & $z_{p1}$ & 19.54 & 0.04 \\
		56488.4 & -24.2	 & $z_{p1}$ & 19.55 & 0.03 \\
		56490.5 & -22.1	 & $z_{p1}$ & 19.48 & 0.05 \\
		56492.5 & -20.1	 & $z_{p1}$ & 19.54 & 0.04 \\
		56506.6 & -6.0	 & $z_{p1}$ & 19.53 & 0.04 \\
		56511.6 & -1.0	 & $z_{p1}$ & 19.47 & 0.04 \\
		56517.4 & 4.8	 & $z_{p1}$ & 19.49 & 0.04 \\
		56519.6 & 7.0    & $z_{p1}$ & 19.60 & 0.04 \\
		56521.6 & 9.0  	 & $z_{p1}$ & 19.65 & 0.04 \\
		56530.4 & 17.8	 & $z_{p1}$ & 19.81 & 0.04 \\
		56536.5 & 23.9	 & $z_{p1}$ & 19.84 & 0.05 \\
		56539.4 & 26.8	 & $z_{p1}$ & 19.91 & 0.05 \\
		56541.3 & 28.7	 & $z_{p1}$ & 19.79 & 0.05 \\
		56544.3 & 31.7	 & $z_{p1}$ & 19.87 & 0.06 \\
		56549.3 & 36.7	 & $z_{p1}$ & 19.90 & 0.06 \\
		56557.5 & 44.9	 & $z_{p1}$ & 20.07 & 0.06 \\
		56568.3 & 55.7	 & $z_{p1}$ & 20.35 & 0.14 \\
		56570.3 & 57.7	 & $z_{p1}$ & 20.03 & 0.06 \\
		56584.3 & 71.7	 & $z_{p1}$ & 19.92 & 0.05 \\
		56589.3 & 76.7	 & $z_{p1}$ & 19.94 & 0.05 \\
		56592.3 & 79.7	 & $z_{p1}$ & 19.87 & 0.05 \\
		56595.2 & 82.6	 & $z_{p1}$ & 19.84 & 0.05 \\
		56600.3 & 87.7	 & $z_{p1}$ & 19.81 & 0.05 \\
		56613.3 & 100.7	 & $z_{p1}$ & 19.87 & 0.05 \\
		56628.2 & 115.6	 & $z_{p1}$ & 20.03 & 0.06 \\
		56266.3 & -246.0 & $y_{p1}$ & (21.53) &         \\
		56434.6 & -78.0  & $y_{p1}$ & 20.63 & 0.17 \\
		56436.6 & -76.0  & $y_{p1}$ & 20.42 & 0.10 \\
		56437.6 & -75.0  & $y_{p1}$ & 20.73 & 0.29 \\
		56454.6 & -58.0  & $y_{p1}$ & 19.70 & 0.11 \\
		56467.5 & -45.0  & $y_{p1}$ & 19.82 & 0.04 \\
		56484.5 & -28.0  & $y_{p1}$ & 19.34 & 0.05 \\
		56493.6 & -19.0  & $y_{p1}$ & 19.50 & 0.03 \\
		56494.6 & -18.0  & $y_{p1}$ & 19.58 & 0.03 \\
		56511.5 & -1.0	  & $y_{p1}$ & 19.37 & 0.03 \\
		56517.4 & 5.0	  & $y_{p1}$ & 19.32 & 0.06 \\
		56522.6 & 10.0	  & $y_{p1}$ & 19.66 & 0.04 \\
		56532.4 & 20.0	  & $y_{p1}$ & 19.66 & 0.04 \\
		56556.4 & 44.0	  & $y_{p1}$ & 19.73 & 0.03 \\
		56559.3 & 47.0	  & $y_{p1}$ & 19.94 & 0.09 \\
		56562.3 & 50.0	  & $y_{p1}$ & 19.88 & 0.11 \\
		56563.2 & 51.0	  & $y_{p1}$ & 19.89 & 0.04 \\
		56566.5 & 54.0	  & $y_{p1}$ & 20.04 & 0.17 \\
		56567.3 & 55.0	  & $y_{p1}$ & 19.77 & 0.14 \\
		56569.2 & 57.0	  & $y_{p1}$ & 20.03 & 0.19 \\
		56573.3 & 61.0	  & $y_{p1}$ & 20.00 & 0.08 \\
		56574.2 & 62.0	  & $y_{p1}$ & 19.89 & 0.19 \\
		56584.4 & 72.0	  & $y_{p1}$ & 19.72 & 0.04 \\
		56585.3 & 73.0	  & $y_{p1}$ & 19.96 & 0.07 \\
        56586.3 & 74.0	  & $y_{p1}$ & 20.15 & 0.05 \\
		\bottomrule
	\end{tabular}
\end{table}

\begin{table}
%\contcaption{}
\renewcommand\thetable{1}
\centering \caption{\textit{continued}}
%\label{tab:continued}
\begin{tabular}{clcccc}
		\toprule
      Data of Observation & \multicolumn{1}{l}{Epoch} & Filter & Mag  & Mag \\
        (MJD) & \multicolumn{1}{c}{(days)} & & (observed) &  Uncertainty\\
        \midrule
        
        56613.3 & 101.0  & $y_{p1}$ & 19.79 & 0.04 \\
		56616.2 & 104.0  & $y_{p1}$ & 19.88 & 0.08 \\
		56620.2 & 108.0  & $y_{p1}$ & 19.80 & 0.04 \\
		56626.2 & 114.0  & $y_{p1}$ & 19.70 & 0.17 \\
		56638.2 & 126.0  & $y_{p1}$ & 19.77 & 0.04 \\
		\bottomrule
		\
	\end{tabular}
	\begin{flushleft}
    \textbf{Note:} Magnitudes provided here are not reddening corrected and the epochs are relative to the peak in the rest frame. $3\sigma$ upper limit values are represented in parentheses.
    \end{flushleft}
\end{table}

% Alternatively you could enter them by hand, like this:
% This method is tedious and prone to error if you have lots of references
%\begin{thebibliography}{99}
%\bibitem[\protect\citeauthoryear{Author}{2012}]{Author2012}
%Author A.~N., 2013, Journal of Improbable Astronomy, 1, 1
%\bibitem[\protect\citeauthoryear{Others}{2013}]{Others2013}
%Others S., 2012, Journal of Interesting Stuff, 17, 198
%\end{thebibliography}

%%%%%%%%%%%%%%%%%%%%%%%%%%%%%%%%%%%%%%%%%%%%%%%%%%

%%%%%%%%%%%%%%%%% APPENDICES %%%%%%%%%%%%%%%%%%%%%
%
%\appendix
%
%\section{Some extra material}
%
%If you want to present additional material which would interrupt the flow of the main paper,
%it can be placed in an Appendix which appears after the list of references.

%%%%%%%%%%%%%%%%%%%%%%%%%%%%%%%%%%%%%%%%%%%%%%%%%%

% Don't change these lines
\bsp	% typesetting comment
\label{lastpage}
\end{document}